\documentclass[english]{article}
\usepackage[T1]{fontenc}
\usepackage[latin9]{inputenc}
\usepackage{geometry}
\usepackage{color}
\usepackage{verbatim}
\usepackage{amsmath}
\usepackage{amssymb}
\usepackage{graphicx}
\usepackage{setspace}
\usepackage{authblk}
\makeatletter

\providecommand{\tabularnewline}{\\}
\date{}

\@ifundefined{showcaptionsetup}{}{%
 \PassOptionsToPackage{caption=false}{subfig}}
\usepackage{subfig}
\makeatother

\usepackage{babel}
\begin{document}
\title{\textcolor{black}{A solution to the permalloy problem}}
\author[1]{Ananya Renuka Balakrishna}
\author[2]{Richard D. James}
\affil[1]{\small{Aerospace and Mechanical Engineering, University of Southern California, Los Angeles CA 90089}}
\affil[2]{\small{Aerospace Engineering and Mechanics, University of Minnesota, Minneapolis MN 55455}}
\maketitle
\begin{abstract}
We propose a solution to the longstanding permalloy problem---why the particular composition of permalloy, $\mathrm{Fe_{21.5}Ni_{78.5}}$, achieves a dramatic drop in hysteresis, while its material constants show no obvious signal of this behavior. We use our recently developed coercivity tool to show that a delicate balance between local instabilities and magnetic material constants are necessary to explain the dramatic drop of hysteresis at 78.5$\%$ Ni. Our findings are in agreement with the permalloy experiments and, more broadly, provide theoretical guidance for the discovery of novel low hysteresis magnetic alloys.

\end{abstract}

\section*{\textcolor{black}{Introduction}}

\textcolor{black}{}%

\begin{figure}
\begin{centering}
\textcolor{black}{\includegraphics[width=\textwidth]{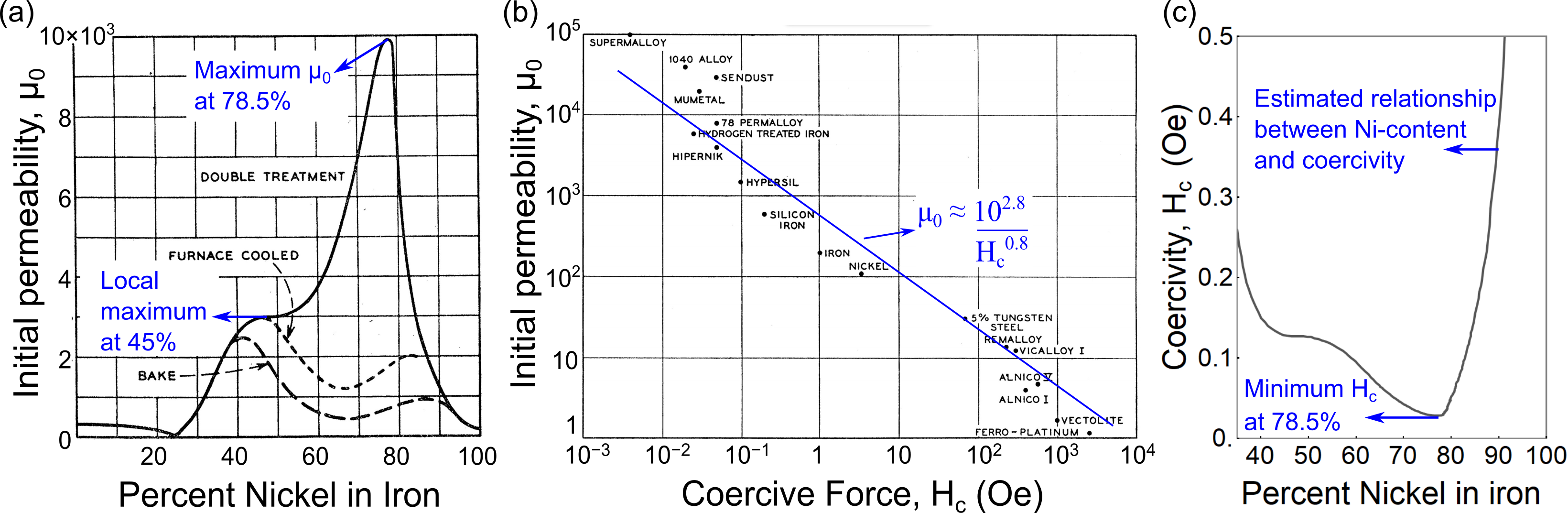}}
\par\end{centering}
\textcolor{black}{\caption{\label{IntroductionPlots}(a) A permeability plot for Fe-Ni alloys as a function of Ni-content. The highest initial permeability is achieved at 78.5$\%$ Ni-content, and a second peak is observed at 45$\%$ Ni-content.  (b) The initial permeability correlates inversely, $\mu_0 \approx 10^{2.8}/\mathrm{H_c}^{0.8}$ with coercivity in magnetic alloys. (c) Using subfigures a-b, we estimate the relationship between Ni-content and coercivity in magnetic alloys. We note that the minimum coercivity is at 78.5$\%$ Ni-content. \textcolor{black}{ Subfigures a-b are reprinted with permission from Bozorth, R. M, Reviews of Modern Physics, 25(1), 42, 1953 and Kittel, C. Reviews of Modern Physics, 21(4), 541, 1949, respectively. Copyright (1953, 1949) by the American Physical Society.}}}
\vspace{-5mm}
\end{figure}

In the early 20th century, an Fe-Ni alloy with unusually low coercivity was discovered at  Bell Laboratories \cite{PermalloyDiscovery}. This magnetic alloy with precisely $78.5\%$ Ni, now known as  permalloy, demonstrated a drastic lowered  hysteresis, quantified by the value of coercivity,  relative to nearby alloys, see Fig.~\ref{IntroductionPlots}(a-c). Several researchers attribute the dramatic decrease in hysteresis at $78.5\%$ Ni to its small anisotropy constant $\kappa_1$ -- a material constant that quantifies the difficulty of rotating the magnetization away from certain preferred crystallographic axes \cite{BozorthFerromagnetism}. However, a closer examination of the binary Fe-Ni alloys, containing $35\%-100\%$ Ni, shows several peculiarities in behavior that contradict our current understanding of the origins of magnetic hysteresis. For example, the anisotropy constant of the FeNi alloy is zero at $75\%$ Ni; however, there is not even a local minimum of the coercivity vs.~composition at this Ni-content. However, at $78.5\%$ Ni, where the anisotropy constant is clearly not zero, the coercivity is minimized \cite{BozorthPermalloyProblem} (see supplementary information).

Besides the anisotropy constant, researchers have suspected that the magnetostriction constants,\footnote{These material constants relate the preferred strains corresponding to a given magnetization.} $\lambda_{111}$ and $\lambda_{100}$, play some role
in lowering magnetic hysteresis \cite{BeckerDoring,Kittel,Lewis}. 
The potential influence of $\lambda_{100}$ is supported by the presence of
the second permalloy composition at $45\%$ Ni, where the coercivity
vs.~composition shows a diffuse local minimum, not nearly as sharp as in $78.5\%$ Ni, but still clearly noticeable, see Fig.~\ref{IntroductionPlots}(a,c).  The values of anisotropy constant $\kappa_1$ and magnetostriction constant $\lambda_{111}$ are far
from zero at $45\%$ Ni, but the magnetostriction constant $\lambda_{100}=0$ vanishes precisely at this composition. Following
this line of argument, we
would expect to see a lowering of hysteresis at $80\%$ and $83\%$
Ni, at which the magnetostriction constants $\lambda_{111}$
and $\lambda_{100}$ are zero, respectively.  However, there is
not even a discernible local minimum of coercivity vs.~composition
at these 
compositions. By contrast, as mentioned above, the magnetic hysteresis is minimum at $78.5\%$ Ni at which neither of the magnetostriction constants nor the anisotropy constant are zero.
This collection of apparently contradictory facts is known as the 
{\it permalloy problem} \cite{BozorthPermalloyProblem}. 

Processing also has an important influence on hysteresis. For example, slow  cooling can raise the composition of lowest hysteresis to higher Ni content \cite{BozorthFerromagnetism}.  Typically, the effect of a thermo-mechanical treatment in the Fe-Ni system is to induce the formation of precipitates.  This in turn has two main effects: (a) it leads to the preferential deposition of one element into  precipitates, which have a very different composition than the matrix. This precipitation causes a departure of the matrix composition from the nominal composition; and (b) it develops residual stress due to the geometric incompatibility of the precipitate and the matrix. This geometric incompatibility induces coherency stresses that affect magnetic properties via magnetostriction \cite{precipitation-1,precipitation-2}. In certain systems more disruptive processing treatments, such as rapid solidification leading to nanocrystalline materials, is also a productive route toward extremely low hysteresis materials \cite{Willard,McHenry,flash-milling}. 

Researchers have attempted to resolve the permalloy problem for many years \cite{BozorthPermalloyProblem,BeckerDoring,Lewis}. 
Domain theory has been used to explain the ease of domain rotation and domain wall movement at small values of anisotropy and magnetostriction constants \cite{BeckerDoring, Kittel}. These calculations make certain assumptions on microstructures and domain structure, and do not lead  to the permalloy composition. In another study \cite{Lewis}, the criterion $|(\lambda_{100}-\lambda_{111})\sigma|=|\kappa_1|$ between magnetostriction and anisotropy constants (in the presence of residual stresses $\sigma$)  is proposed as governing 
coercivity.  While we agree with the importance of accounting for $\lambda_{100}, \lambda_{111}, \sigma \thinspace \mathrm{and} \thinspace \kappa_1$ in predicting coercivity, we do not see a fundamental theoretical or experimental basis for this criterion.
In addition, since residual stress is a tensor varying with position in a heterogeneous solid, it is not clear to us how to use this criterion. Several recent studies examine effect of grain orientation and crystallographic texture on coercivity in FeNiMo alloys \cite{exp1,exp2}, however these studies do not address how fundamental material constants interact with defects and residual stress to lower hysteresis in this system.

In our paper, while recognizing that processing 
can be highly influential, we explore the hypothesis that there is a relation between  hysteresis and fundamental material constants in the FeNi system. Doing so, we shed light on the permalloy problem.

Recently, we developed a coercivity tool based on the micromagnetics theory \cite{ARBJames}. Our tool differs from other theoretical methods that predict magnetic coercivity \cite{RecentModels,ModelGPU}, in two ways: (a) First, we account for magnetoelastic interactions that have been neglected in most prior studies. (b) Second, we introduce an optimized localized disturbance in the form of a spike-domain microstructure that grows during magnetization reversal.  By introducing the localized disturbance, we capture the delicate balance between magnetic material constants that govern hysteresis, and predict coercivity with greater accuracy than other methods based on linear stability analysis. In this work, we apply this coercivity tool to provide insight into the permalloy problem.

In this paper, we first provide an overview of the micromagnetics theory used in our coercivity tool. We then apply this tool in two studies on iron-nickel alloys: In Study 1, we test the hypothesis that a specific combination of magnetoelastic constants (with residual stress) and anisotropy constants is necessary to lower the coercivity at the permalloy composition. Here, we compute coercivity in iron-nickel alloys in two cases, (a) by accounting for magnetostrictive effects and (b) by ignoring all magnetoelastic energy contributions. In Study 2, we test the hypothesis that neither defect geometry nor defect orientation significantly affect the combination of material constants at which the lowest coercivity is achieved. Finally, we compare the magnetic coercivity predicted by the present study with previously proposed ideas \cite{BeckerDoring, Kittel}. Our simulations show that the lowest coercivity is attained at $78.5\%$ Ni-content when magnetoelastic energy is accounted in the model, and provides insight into minimum coercivity values of other Fe-Ni alloys.

\begin{figure}
\begin{centering}
\textcolor{black}{\includegraphics[width=\textwidth]{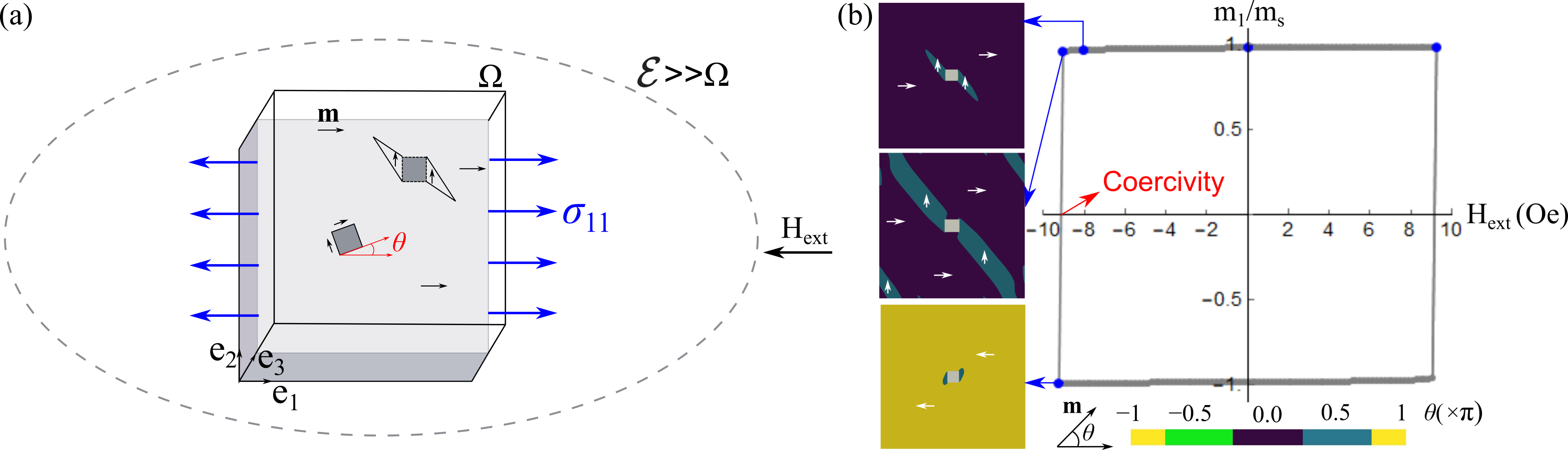}}
\par\end{centering}
\textcolor{black}{\caption{\label{SchematicEllipsoid}(a) A 3D computational domain $\Omega$ with nonmagnetic inclusions. This computational domain is much smaller than the size of the ellipsoid $\mathcal{E}$. Other parameters modeled in the present study, such as defect orientation $\theta$ and applied stresses $\sigma_{11}$ and external field $\mathrm{\mathbf{H}_{ext}}$, are schematically illustrated. (b) A representative calculation of magnetic coercivity. We apply a large external field H$_\mathrm{ext}$ in the direction of magnetization and decrease its value incrementally. At a critical field strength, known as coercivity, the magnetization reverses. The inset figures show microstructures during magnetization reversal and the color bar represents the orientation of magnetization.}
}
\vspace{-5mm}
\end{figure}

\section*{\textcolor{black}{Theory and Methods}}

\textcolor{black}{}%

\textcolor{black}{Our coercivity tool is based on micromagnetics with magnetostriction \cite{BrownMicromagnetics,BrownMagnetoelastic}.
The total free energy is a functional of magnetization $\mathbf{m}$,
$|\mathbf m| = m_s$,
and strain $\mathbf{E} = \frac{1}{2}(\nabla \mathbf u + \nabla \mathbf u^T)$ given by}

\begin{align}
{\Psi} & =\int_{\mathcal{E}}\{\nabla\mathbf{m}\cdot\mathrm{A\nabla\mathbf{m}+\kappa_{1}(\mathrm{m_{1}^{2}m_{2}^{2}}+\mathrm{m_{2}^{2}m_{3}^{2}}+\mathrm{m_{3}^{2}m_{1}^{2}})}+\frac{1}{2}[\mathbf{E}-\mathbf{E_{0}\mathrm{(\mathbf{m})}}]\cdot\mathbb{C}[\mathbf{E}-\mathbf{E_{0}\mathrm{(\mathbf{m})}}]\nonumber\\
 & -\mathbf{\sigma_\mathrm{ext}}\cdot\mathbf{E}-\mu_{0}\mathbf{H_{\mathrm{ext}}\cdot m}\}\mathrm{d\mathbf{x}} +\int_{\mathbb{R}^{3}}\mu_{0}\left|\mathbf{H}_\mathrm{d}\right|^{2}\mathrm{d\mathbf{x}},\label{MicromagneticsEnergy}
 \end{align}
and $\mathbf{\sigma}_{\mathrm{ext}}, \mathbf{H}_{\mathrm{ext}}$ are the  (constant) applied stress and magnetic field, respectively. Here, the anisotropy energy
$\kappa_{1}(\mathrm{m_{1}^{2}m_{2}^{2}}+\mathrm{m_{2}^{2}m_{3}^{2}}+\mathrm{m_{3}^{2}m_{1}^{2}})$
penalizes rotation of the magnetization away from the easy
axes, the elastic energy $\frac{1}{2}[\mathbf{E}-\mathbf{E_{0}\mathrm{(\mathbf{m})}}]\cdot\mathbb{C}[\mathbf{E}-\mathbf{E_{0}\mathrm{(\mathbf{m})}}]$
penalizes mechanical deformation away from the preferred strain, defined by
\begin{align}
\mathbf{E_{0}}(\mathbf{m}) & =\mathrm{\frac{3}{2}\left[\begin{array}{ccc}
\lambda_{100}(\mathit{m}_{1}^{2}-\frac{1}{3}) & \lambda_{111}\mathit{m}_{1}\mathit{m}_{2} & \lambda_{111}\mathit{m}_{1}\mathit{m}_{3}\\
 & \lambda_{100}(\mathit{m}_{2}^{2}-\frac{1}{3}) & \lambda_{111}\mathit{m}_{2}\mathit{m}_{3}\\
\mathit{symm.} &  & \lambda_{100}(\mathit{m}_{3}^{2}-\frac{1}{3})
\end{array}\right]}\label{eq:PreferredStrain}.
\end{align}
Finally, the vector field $\mathbf{H}_\mathrm{d}$ is the stray field generated by the magnetization distribution, 
and the magnetostatic energy $\mu_{0}\left|\mathbf{H}_\mathrm{d}\right|^{2}$ is the energy required to assemble a collection of elementary magnetic dipoles into the given magnetization distribution \cite{james1994}. 
The stray field $\mathbf{H}_\mathrm{d} = - \nabla \zeta_{\mathbf m} $ is obtained by solving the magnetostatic equation $ \nabla \cdot (-\nabla \zeta_{\mathbf m} + 
\mathbf m)=0$ on all of space.   Here, we note that
in this model the five material constants $\mathrm{A}, \kappa_1, \lambda_{100}, \lambda_{111}, m_s$ together with the elastic modulus tensor $\mathbb{C}$ (depending on the three moduli $c_{11}, c_{12}, c_{44}$) and applied field and stress determine the form
of the micromagnetic energy.
See \cite{ARBJames} for further explanation.

\textcolor{black}{We compute the evolution of the magnetization 
using a local energy minimization technique, based on the generalized Landau-Lifshitz-Ginzburg
equation \cite{LLG}:}

\begin{align}
\frac{\partial\mathbf{m}}{\partial t}=-\gamma\mathbf{m}\times\mathcal{H}-\frac{\gamma\alpha}{m_{s}}\mathbf{m}\times(\mathbf{m}\times\mathcal{H}).\label{LLG}
\end{align}

\textcolor{black}{Here, $\mathcal{H}=-\frac{\delta\Psi}{\delta\mathbf{m}}$
is the effective field, $\gamma$ is the gyromagnetic ratio, and $\alpha$
is the damping constant. We numerically solve Eq.~\ref{LLG} using the Gauss
Siedel Projection method \cite{GSPM}, and identify equilibrium states
when the magnetization evolution converges, $\left|\mathbf{m^{\mathrm{n+1}}}-\mathrm{\mathbf{m}^{n}}\right|<10^{-9}$.
At each iteration we compute the magnetostatic field $\mathbf{H_{d}}=-\nabla\zeta_{\mathbf{m}}$
and the strain  $\mathbf{E}$ by solving their respective equilibrium
equations:}

\textcolor{black}{
\begin{align}
\nabla\cdot(\mathbf{\mathbf{-\nabla\zeta_{\mathbf{m}}}+m}) & =0\qquad\mathrm{on}\thinspace\mathbb{R}^{3}\label{MagnetostaticEqbm}\\
\nabla\cdot\mathbb{C}(\mathbf{E-E}_{0}) & =0.\label{MechanicalEqbm}
\end{align}
}

\textcolor{black}{The magnetostatic equilibrium condition arises from
the Maxwell equations, namely $\nabla\times\mathbf{H_{d}}=0\to\mathbf{H_{d}}=-\nabla\zeta_{\mathbf{m}}$
and $\nabla\cdot\mathbf{B}=\nabla\cdot(\mathbf{H_{d}+m})=0$. We compute the magnetostatic and mechanical equilibrium conditions in Eqs.~\ref{MagnetostaticEqbm}-\ref{MechanicalEqbm} in Fourier space \cite{FourierChen}. Further details on the numerical method can
be found in Refs. \cite{ARBJames,FourierChen,GSPM}}

\textcolor{black}{In the present work, we calibrate our micromagnetics
model for the FeNi alloy as a function of Ni-content. The
material constants used in our calculations are listed in the supplementary material. We assume an oblate (pancake-like shaped) ellipsoid under a suitably oriented applied field, see 
Fig.~\ref{SchematicEllipsoid}(a).}\footnote{\textcolor{black}{The applied field is directed along the major axis of the magnetic ellipsoid. The easy axes for a magnetic body with anisotropy constants $\kappa_{1}>0$ and $\kappa_{1}<0$ are along $\{100\}$ and $\{111\}$ crystallographic directions, respectively.}}\textcolor{black}{{} This ellipsoid is uniformly magnetized except at the proximity of defects. Using a computational scheme based on the ellipsoid and reciprocal theorems\footnote{In our method the presence of the large ellipsoidal body $\cal E$
is essential as a way of describing the poles on the boundary of the macroscale body without having to simulate the external fields due to these poles. See Ref.~\cite{ARBJames}.}, we model a 3D computational domain $\Omega$ with a nonmagnetic inclusion at its geometric center. The domain and the inclusion are of sizes $64\times64\times24$ and $12\times12\times6$ grid points. The cell size is chosen such that the domain walls span across 3-4 elements. The magnetization  inside the inclusion is held at zero throughout the computation, $\mathrm{\left|\mathbf{m}\right|=\mathrm{0}}$. Outside the defect, we initialize the computation domain with a uniform magnetization $\mathrm{\mathbf{m}=\mathrm{m_{1}}},$ and apply a large external field along the easy axes, $\mathbf{H}_{\mathrm{ext}}>>0$, see Fig.~\ref{SchematicEllipsoid}(b). As we decrease the applied field the spike domain at first grows slowly until a critical field is reached at which the magnetization vector reverses abruptly. 
We identify this value of the applied field with coercivity. We use this approach to predict coercivity of the iron-nickel alloys as a function of Ni-content. }

\section*{\textcolor{black}{Results}}

\begin{figure}
\begin{centering}
\textcolor{black}{\includegraphics[width=0.8\textwidth]{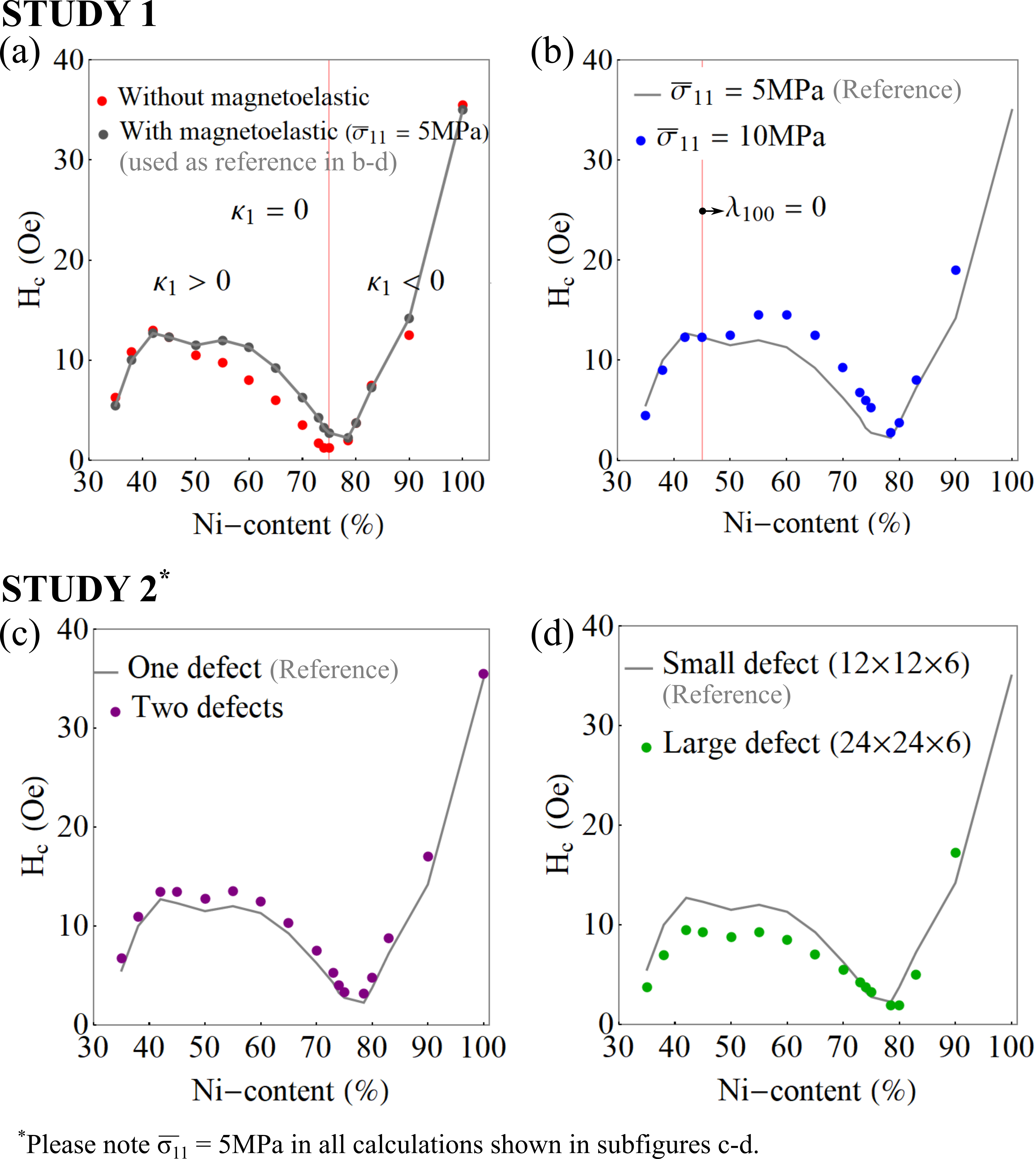}}
\par\end{centering}

\textcolor{black}{\caption{\label{CoercivityPlot}Computed coercivity values in iron-nickel alloys as a function of Ni-content in (a-b) Study 1 and (c-d) Study 2. (a) When both the anisotropy and magnetoelastic energies (including residual stresses $\sigma_\mathrm{11}$ are accounted in the free energy function, the lowest coercivity is predicted at $78.5\%$ Ni-content. When magnetoelastic energy is neglected and only anisotropy energy is accounted in the free energy function, the lowest coercivity is predicted at $75\%$ Ni-content. (b) The minima at $78.5\%$ and $45\%$ Ni-content alloys are more pronounced in iron-nickel alloys under tensile stresses, for example $\sigma_{11} = 10\mathrm{MPa}$. The reference plot corresponds to the coercivity values of FeNi alloys with magnetoelastic interactions and residual stresses $(\lambda_{100}, \lambda_{111}, \sigma_\mathrm{11})$ in subfigure 2(a) (c) Defect orientation and density affect the coercivity values, however, their effect is small. (d) The size of the defect modeled on the computational domain also affects coercivity values in FeNi alloys. }
}
\end{figure}

In Study 1, we test our hypothesis that the magnetoelastic
energy contributions (e.g., magnetostrictive and residual stress), in addition to the anisotropy energy, are necessary to reproduce
the characteristic features of coercivity vs.~composition in iron-nickel
alloys. To test this hypothesis, we compute the coercivity for
iron-nickel alloys in two cases: (a) by accounting for the magnetoelastic
energy with a small residual stress $\sigma_{11}=\mathrm{5MPa}$, and (b) by ignoring all magnetoelastic energy in
Eq.~\ref{MicromagneticsEnergy}. Note that a residual stress of $\sigma_{11}=\mathrm{5MPa}$ corresponds to a uniaxial strain of about 25$\mu\epsilon$ (with typical Young's Modulus of about $200\mathrm{GPa}$) for FeNi alloys. As noted above, these residual stresses naturally arise from precipitation during heat treatment, and from other defects such as the presence of grain boundaries and, in particular, triple junctions. For example a temperature change of 1K causes about $10\mu\epsilon$ strain in iron. Our assumption of $25\mu\epsilon$ residual strains is well within the thermal strain for moderate temperature changes.

Fig.~\ref{CoercivityPlot}(a) shows the coercivity of iron-nickel alloys for the two cases. In line with the experimental evidence, the coercivity is the lowest at $78.5\%$ Ni-content when both magnetoelastic
and anisotropy energy contributions are accounted in the free energy
function. When magnetoelastic energy is neglected (or small), the coercivity is minimum at $75\%$ Ni-content at which $\kappa_{1}=0$. Similarly, our calculations show that the coercive
field has a local minimum at $45\%-50\%$ Ni-content, when both magnetoelastic
and anisotropy energy contributions are included in the model. Furthermore, this minima is more pronounced with increased stresses in the material, see Fig. \ref{CoercivityPlot}(b). Although
the anisotropy constant is large in this composition range $\kappa_{1}\approx10^{3}\mathrm{J/m^{3}}$,
we find that its coercivity is relatively small in this neighborhood.
This is consistent with previous hypotheses by researchers \cite{BozorthPermalloyProblem,RDJ}, who speculated
that the low magnetostriction constant $\lambda_{100}\to0$ in the
$45\%\leq\mathrm{Ni}\leq50\%$ neighborhood causes a local minimum
(or maximum) in the FeNi coercivity (or permeability) plot. When $\lambda_{100}=0$ the strain values are decoupled from the magnetization vector, and consequently the residual stresses do not affect the coercivity values.

\textcolor{black}{In Study 2, we test our hypothesis that neither
defect geometries nor defect densities affect the balance between material
constants at which minimum coercivity is achieved. Fig. \ref{CoercivityPlot}(c-d) shows the effect of defect geometry (such as orientation and shape) and density on the coercivity of the FeNi alloy. Note, these computations were modelled with a residual stress of $\sigma_{11}=\mathrm{5MPa}$.  The magnitude of the coercivity is sensitive to defect geometries and tensile stresses, however,
as hypothesized, for a given defect configuration the minimum coercivity remains at $78.5\%$ Ni-content. The balance between anisotropy
and magnetostriction constants that minimizes the coercivity at
$78.5\%$ Ni-content is not significantly affected by the changes in defect geometry and/or orientation. These results are consistent
with a wider set of variations of defect size and placement presented
in  \cite{ARBJames} (but not for material constants of permalloy).
}%

We next compare our findings with predictions from Domain Theory and Lewis criterion \cite{BeckerDoring,Kittel,Lewis}. We do not include results from linear stability analysis of the single domain state based on micromagnetics, because these vastly over predict the coercivity.\footnote{This over prediction is known as the ``coercivity paradox''.  Our simulations are outside the regime of linear stability analysis.} We compare our results with prior predictions by choosing specific compositions of the iron-nickel alloys (e.g., alloys with 45,75,78.5,80, and 83$\%$ Ni-content) to highlight the zeros of the material constants (see supplementary information): We note that the formulae from Domain Theory $\mu_0|_{\kappa_1} \approx \frac{2m_s^2}{\kappa_1}$ or $\mu_0|_\lambda\approx \frac{5m_s^2}{\lambda^2\mathrm{E}}$ predicts large permeabilities $(\to \infty)$ at multiple singularities, however, they fail to identify the highest permeability (lowest coercivity) at the $78.5\%$ Ni-content alloy. For example, Domain Theory predicts the highest permeability at $75\%$ Ni-content with $\kappa_1=0$ and at $45\%$ Ni-content with $\lambda_{100}=0$, however, experiments indicate that the highest permeability occurs at 78.5$\%$ Ni-content at which neither $\kappa_1$ nor $\lambda_{100}$ are zero. Similarly, the Lewis criterion $|(\lambda_{100}-\lambda_{111})\sigma|=|\kappa_1|$ is closely satisfied at $75\%$ Ni-content with a value $\sigma=5\mathrm{MPa}$, however, it misses the dramatic drop in coercivity at the $78.5\%$ alloy composition. Furthermore, it is unclear on how to choose the residual stress $\sigma$ in the criterion; we keep it at $\sigma=5\mathrm{MPa}$
to reflect the fact that the measured values were obtained from alloys with nominally the same heat treatment. By contrast, our coercivity tool predicts the highest permeability (or lowest coercivity) at $78.5\%$ Ni-content, see Fig.~\ref{CoercivityPlot}(a-b). Furthermore, it shows a local minimum in coercivity at $45\%$ Ni, and provides insight into coercivities at other alloy compositions with zero material constants. Overall, the prior criteria or formulae that explain the low coercivity at the permalloy composition are not general, and do not fully explain the singularities at other material compositions. These results suggest that  our approach based on non-linear stability analysis at the shoulder of the hysteresis loop together with a delicate balance of magnetic material constants is a potential way forward to reliably predict coercivities.

\section*{\textcolor{black}{Discussion}}

\textcolor{black}{In summary, our findings on coercivity as a
function of Ni-content in the binary FeNi alloys show that the lowest
coercivity is attained at $78.5\%$ Ni-content. This was the case
in Study 1 when both anisotropy and magnetoelastic energy contributions
were included in the calculations. In Study 2, although tensile stresses
and defect geometries affected the coercivity values, the minimum
coercivity was still observed at $78.5\%$ Ni-content. Furthermore, our predictions on coercivity as a function of Ni-content are more accurate than prior criterion based on domain theory, and provides insight into the permalloy problem. Below, we discuss
some limiting conditions on our results and then highlight the key features of our work.}

Two features of this work limit the conclusions
we can draw about the permalloy problem. First, we assume a single
crystalline ellipsoid body, which does not contain grain boundaries,
sharp edges and other imperfections, which are commonly found in bulk
magnetic materials. This assumption possibly contributes to about an order of magnitude difference between our predicted coercivity values and those reported from experiments. Second, we model the local disturbance (spike-domain) explicitly by defining a potent defect on the computational domain. Whether developing a more fundamental theory of nucleation in the Calculus of Variations without explicitly defining the nucleus would provide further insights into magnetic coercivity is an open question. With these reservations in mind we next discuss our findings on the permalloy problem.

The key feature of our work is that we show a delicate balance between material constants, $\kappa_1,\lambda_{100},\lambda_{111}$, is necessary for the low coercivity at $78.5\%$ Ni-content. This finding contrasts with some reports in the literature in which the small anisotropy constant $\kappa_{1}\to0$ is considered to be the $\mathit{only}$ factor responsible for lowering hysteresis. Although $\kappa_{1}=0$ lowers the coercivity (e.g., at $75\%$ Ni-content), we find that  magnetostriction plays an important role in governing magnetic hysteresis. Furthermore, our findings are consistent with previous hypotheses that the zero magnetostriction constants---along the easy axes---at $45\%$ and $80\%$ Ni-contents help lower the coercivity values. Despite the large anisotropy constants at these compositions, we find that their coercivities are relatively small in their local Ni-content neighbourhoods.

Another feature of our work is that we propose a theoretical method to compute coercivity, on the shoulder of the hysteresis loop, by accounting for localized instabilities (spike domain). This approach of computing non-linear stability analysis helped us to elucidate the role of magnetic material constants on coercivity. We believe that this method helped us to predict coercivities more accurately when compared to other methods that are based on linearization. In our future work, we use our tool to compute coercivity in a broader material parameter space \cite{HeatMap}, and seek to identify the fundamental relation between the material constants that governs hysteresis.

\section*{\textcolor{black}{Conclusion}}

To conclude, the present findings contribute
to a more nuanced understanding of how material constants, such as
anisotropy and magnetostriction constants, affect magnetic hysteresis.
Specifically, magnetoelastic interactions have been regarded to play
a negligible role in lowering the coercivity. Given the
current findings, we quantitatively demonstrate that the interplay
between anisotropy energy, magnetoelastic energy, and localized disturbance (spike domain) is necessary to lower magnetic hysteresis. Our theoretical
model serves as a design tool to discover novel combinations of material
constants, which lower the coercivity in magnetic alloys.

\vspace{2mm}
\noindent {\bf Acknowledgment}. The authors acknowledge the Minnesota Supercomputing Institute at the University of Minnesota (Dr. David Porter), and the High-Performance Computing Center at the University of Southern California for providing resources that contributed to the research results reported within this paper. The authors thank NSF (DMREF-1629026) and ONR (N00014-18-1-2766) for partial support of this work.  R.D.J and A.R.B, respectively, also acknowledge the support of a Vannevar Bush Faculty Fellowship and a Provost Assistant Professor Fellowship.

\newpage
\section*{\textcolor{black}{Supplementary information}}

Table \ref{tab:materialconstant} shows the magnetic material constants, namely the anisotropy constant $\kappa_1$, the magnetostriction constants $\lambda_{100}, \lambda_{111}$, and the saturation magnetization $\mathrm{m_s}$, of FeNi alloys used in this calculation. The elastic moduli are $c_{11} = 240\mathrm{GPa}$, $c_{12} = 89\mathrm{GPa}$, and $c_{44} = 76\mathrm{GPa}$.

\begin{table}[h]
\begin{centering}
\textcolor{black}{}%
\begin{tabular}{lllll}
\hline 
\textcolor{black}{$\%$ Ni} & \textcolor{black}{$\kappa_{1}$ (kJ/m$^3$)} & \textcolor{black}{$\lambda_{100}$ ($\times 10^{-6}$)} & \textcolor{black}{$\lambda_{111}$ ($\times 10^{-6}$)} & $\mathrm{m_s} \mathrm{(\times 10^{6} A/m)}$\tabularnewline
\hline 
\hline 
\textcolor{black}{35} & 0.462 & -5.85 & 16.9 & 0.94 \tabularnewline
\textcolor{black}{38} & 0.889 & -7.30 & 25.5 & 1.13 \tabularnewline
\textcolor{black}{42} & 1.140 & -3.89 & 32.3 & 1.23 \tabularnewline
\textcolor{black}{45} & 1.100 & 0.0 & 32.8 & 1.26\tabularnewline
\textcolor{black}{50} & 0.958 & 10.0 & 30.9 & 1.25 \tabularnewline
\textcolor{black}{55} & 0.847 & 20.9  & 26.8 & 1.19 \tabularnewline
\textcolor{black}{60} & 0.701 & 26.2  & 22.2 & 1.15 \tabularnewline
\textcolor{black}{65} & 0.500 & 25.6  & 16.5 & 1.07 \tabularnewline
\textcolor{black}{70} & 0.287 & 22.3  & 10.7 & 0.99 \tabularnewline
\textcolor{black}{73} & 0.142 & 18.9 & 7.13 & 0.94 \tabularnewline
\textcolor{black}{74} & 0.052 & 18.5 & 6.29 & 0.92\tabularnewline
\textcolor{black}{75} & 0.000 & 17.2 & 5.46 & 0.90\tabularnewline
\textcolor{black}{77} & -0.084 & 14.5 & 3.68 & 0.87 \tabularnewline
\textcolor{black}{78.5} & -0.161 & 11.8 & 1.91 & 0.84 \tabularnewline
\textcolor{black}{80} & -0.273 & 8.45 & 0.0 & 0.82 \tabularnewline
\textcolor{black}{83} & -0.520 & 0.0 & -2.68 & 0.76 \tabularnewline
\textcolor{black}{90} & -1.600 & -23.2 & -10.8 & 0.64 \tabularnewline
\textcolor{black}{100} & -5.880  & -53.1 & -26.5 & 0.48\tabularnewline
\hline 
\end{tabular}\\
\par\end{centering}

\caption{\label{tab:materialconstant}List of material constants for the FeNi alloy system \cite{BozorthPermalloyProblem}.}
\end{table}

\end{document}